\documentclass{ws-procs975x65}
\def\beq{\begin{equation}}
\def\eeq{\end{equation}}
\def\x{\mathbf{x}}

\begin{document}

\title{Astrophysics in Strong Electromagnetic Fields and Laboratory Astrophysics}

\author{Sang Pyo Kim$^*$}
\address{Department of Physics, Kunsan National University, Kunsan 54150, Korea\\
Center for Relativistic Laser Science, Institute for Basic Science, Gwangju 61005, Korea\\
and Institute of Theoretical Physics, Chinese Academy of Sciences, Beijing 100190, China\\
$^*$E-mail: sangkim@kunsan.ac.kr}

%\maketitle

\begin{abstract}
Recent observations of gravitational waves from binary mergers of black holes or neutron stars and the rapid development of ultra-intense lasers lead strong field physics to a frontier of new physics in the 21st century. Strong gravity phenomena are most precisely described by general relativity, and lasers that are described by another most precisely tested quantum electrodynamics (QED) can be focused into a tiny area in a short period through the chirped pulse amplification and generate extremely high intensity electromagnetic (EM) fields beyond the conventional methods. It is physically interesting to study QED phenomena in curved spacetimes, in which both strong gravitational and electromagnetic fields play important roles. There are many sources for strong gravitational and electromagnetic fields in the sky or universe, such highly magnetized neutron stars, magnetized black holes, and the early universe. We review quantum field theoretical frameworks for QED both in the Minkowski spacetime and curved spacetimes, in particular, charged black holes and the early universe, and discuss the QED physics in strong EM fields, such as the vacuum polarization and Schwinger pair production and their implications to astrophysics and cosmology.
\end{abstract}

\bodymatter

\section{Introduction}\label{sec1}

Recent observations of gravitational waves, GW150914 from a binary black hole inspiral\cite{Abbott:2016blz} and GW170817 from a binary neutron star inspiral,\cite{TheLIGOScientific:2017qsa} directly confirmed Einstein's general relativity. The precision measurement of gravitational waves will open a new window to test any relativistic theory of gravitation and parts of quantum gravity effects and their remnants. Black holes and neutron stars are the sources for strong gravitational fields whose effects can be observed even at cosmic distances. Another source for strong gravitational fields is the early universe in which the curvature effect of gravity, classical or quantum, is not negligible. Neutron stars, in particular, magnetars (highly magnetized neutron stars) provide the most intense magnetic fields in the universe (for a review and references, see Ref. \citenum{Kaspi:2017fwg}), possibly, except for exotic objects, such as cosmic strings and the very early universe. The magnetic fields of magnetars go by order of two or more beyond the critical field $B_c = 4.41 \times 10^{13}\, \mathrm{G}$ of the lowest Landau energy equaling the rest mass of the electron. The quantum nature of matter in neutron stars and magnetars entirely differs from that of ordinary matter in weak electromagnetic (EM) fields in laboratory.\cite{Meszaros:1992} The vacuum birefringence, a vacuum polarization effect, was recently observed by measuring optical spectrum from a neutron star.\cite{Mignani:2016fwz}
The chirped pulse amplification (CPA) technology, however, has overcome obstacles for amplifying lasers beyond the breakdown of optical materials,\cite{Strickland:1985} and by using flying mirrors, can generate EM pulses beyond the critical field for electron-positron pairs.

In this review, we survey the physics and theory underling phenomena in strong EM fields in astrophysics and laboratory experiments, and epitomize quantum electrodynamics (QED) in strong EM fields and/or curved spacetimes as well as the Minkowski spacetime. It has been well known for long that a strong EM field can make the vacuum polarized due to the interaction of the photons with virtual electrons from the Dirac sea,\cite{Heisenberg:1935qt} and that a strong electric field even creates electron-positron pairs, the so-called (Sauter-)Schwinger mechanism, which can be explained by quantum tunneling through a titled barrier by the electric potential.\cite{Sauter:1932ab,Schwinger:1951nm} Schwinger pair production is the most efficient mechanism that produces a pair per unit Compton volume and per unit Compton time at the cost of the electric field energy and thus spontaneously breaks down the vacuum when the electric field reaches the critical strength of the electrostatic potential energy across one Compton wavelength equaling to the rest mass of electron.
The vacuum polarization and Schwinger mechanism is genuinely nonperturbative quantum effects in background gauge fields of the Maxwell theory. On the other hand, black holes also emit all species of particles according to the spin statistics of the emitted particles, known as Hawking radiation,\cite{Hawking:1974sw} as a consequence of pair production near their event horizons due to vacuum fluctuations.\cite{Parikh:1999mf} Hawking radiation has been studied for various black holes in general relativity and modified gravity theories.

The purpose of this review is to present an overview of nonperturbative quantum effects in strong EM fields and/or in curved spacetimes as summarized in Fig.~1 and then to explore a possibility of laboratory astrophysics. Ruffini, Vereshchagin and Xue reviewed astrophysics and laboratory experiments in strong field QED,\cite{Ruffini:2009hg} to which this article is not only complementary but the recent progress and some critical arguments on controversial issues are added. In classical theory, a few solutions of the Einstein-Maxwell equations have been found: electrically and/or magnetically charged black holes and the Bonnor-Melvin universe in a uniform magnetic field, for instance, are most widely studied solutions (for review and references, see Ref.~\citenum{Stephani:2003}). To the best knowledge of the author, however, no solution has been found yet for the Einstein equation equated to the stress-energy tensor of the Maxwell theory together with QED effective action at the one-loop level. Hence, most of studies on this topic assume that the backreaction of the stress-energy tensor from QED action is not significant enough to modify the spacetimes of the Einstein-Maxwell theory or Einstein theory. The theoretical framework is quantum field theory for charged scalar or fermion fields: the Klein-Gordon or Dirac equation in background EM fields in curved spacetimes. As shown in Fig.~1, the most well known spacetimes are charged black holes, such as the Reissner-Nordstr\"om (RN) black hole or the Kerr-Newman (KN) black hole or both electrically and magnetically charged RN and KN black holes, and expanding spacetimes, such as the dS space or the Friedmann-Robertson-Walker (FRW) universe. The maximally symmetric (Anti-)de Sitter $(\mathrm{(A)dS})$ space separates the field equations by conserved charges from the symmetry and thereby defines the in- and out-vacua in the asymptotic regions.

\begin{figure}[t]
\begin{center}
\includegraphics[width=4.75in]{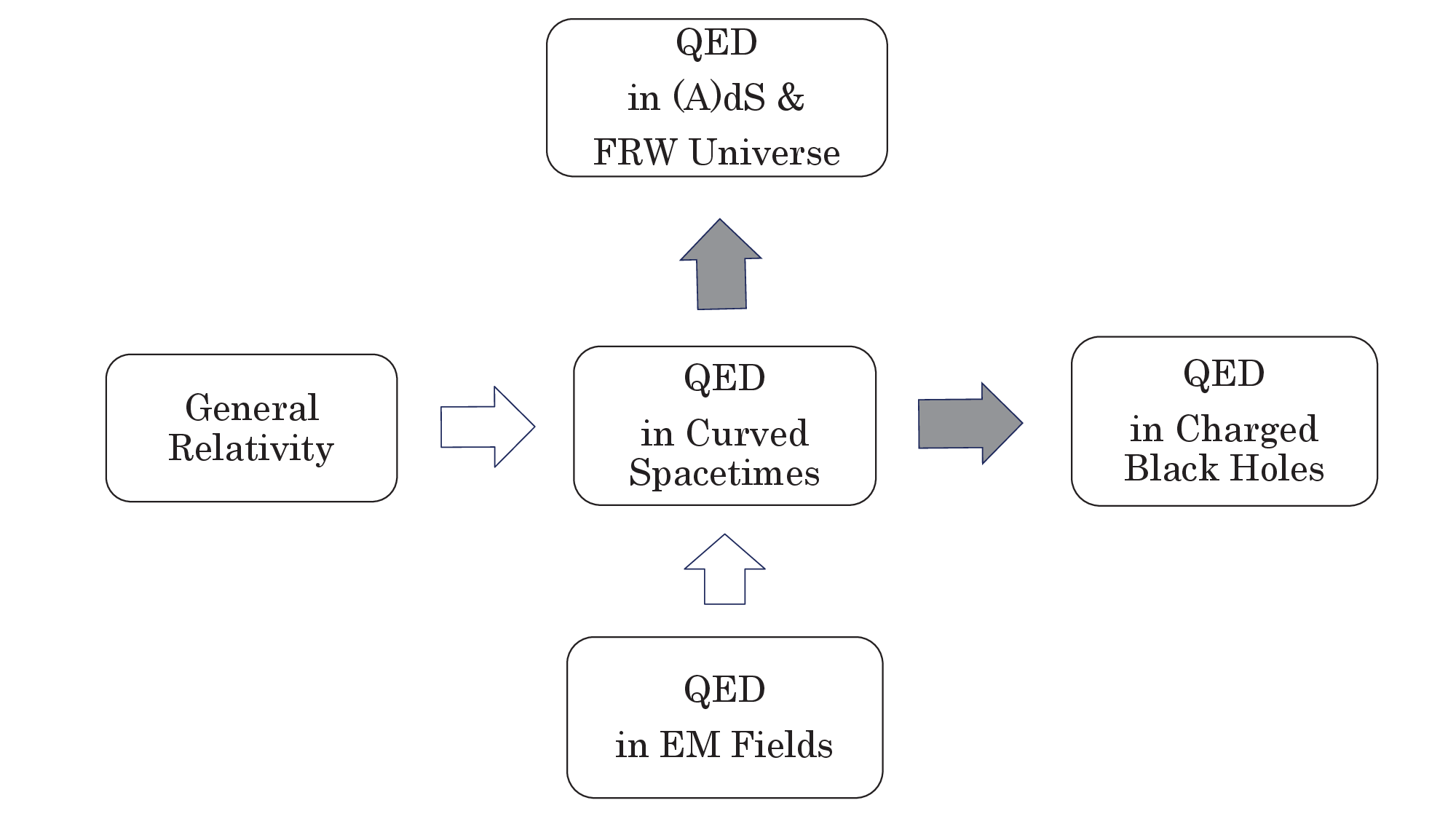}
\end{center}
\caption{The diagram summarizes the scope of this review: QED and Schwinger mechanism in charged black holes and $\mathrm{(A)dS}$ spaces.}
\label{aba:fig1}
\end{figure}

There is vast literature on related topics but many important papers will not be referred to when they are not directly relevant to this review. Thus, only those papers related to the schematic shown Fig.~1 will be referred to. First of all, the role of rotating black holes, electrically charged or with surrounding magnetic fields cannot be overemphasized in astrophysical phenomena. Blanford and Znajek proposed a central mechanism for gamma rays bursts (GRBs), in which magnetized rotating black holes power jets by mining the rotational energy through magnetic fields.\cite{Blandford:1977} Damour and Ruffini studied the Schwinger mechanism in KN black holes and discussed a possibility of astrophysical source for GRBs.\cite{Damour:1974qv} The idea has since then been elaborated to the dyadosphere or dyadotorus model for GRBs (for comprehensive review and references, see Ref.~\citenum{Ruffini:2009hg}). Second, the recent discovery of cosmic scale magnetic fields has motivated investigations of QED in strong EM fields in the dS space or the early universe since the origin of the cosmic magnetic fields, the so-called magnetogenesis, has not been completely revealed yet (for review and references, see Ref.~\citenum{Durrer:2013pga}). Third, the CPA technology has boosted attempts for extremely high intensity of pulsed EM fields and in the near future the intensity of EM fields is likely to reach the strong field QED regime for Schwinger pair production and the vacuum polarization effects etc. Then, a new window for laboratory astrophysics will open for observing some processes of astrophysics in strong EM fields.

The organization of this paper is as follows. In Sec.~\ref{sec2}, we briefly review the approaches to QED effective actions in strong EM fields in the Minkowski spacetime. The in-out formalism based on the Schwinger variational principle and realized for exact one-loop effective actions through the gamma-function regularization explicitly gives QED actions not only in the Minkowski spacetime but also in curved spacetimes. In Sec.~\ref{sec3}, QED is studied in the dS space and the infrared(IR)-hyperconductivity is discussed in connection to the primordial magnetogenesis of cosmological magnetic fields in the early universe. The key concept for the leading Schwinger formula in the $\mathrm{(A)dS}_2$ space is the effective temperature for charges due to the curvature of spacetime as well as the acceleration of charges in the electric field. The effective temperature is universal because the near-horizon geometry of near-extremal black holes has a warping geometry of
 $\mathrm{AdS}_2$. In Sec.~\ref{sec4}, the Schwinger mechanism for charge emission from charged black holes is reviewed and explained in terms of the effective temperature. In Sec.~\ref{sec5}, we review QED physics and recent discoveries in astrophysics, and discuss possible observations in the future. In Sec.~\ref{sec6}, we summarize the recent development in QED in an expanding universe and possible primordial magnetogenesis scenarios.  In Sec.~\ref{sec7}, we propose laboratory astrophysics using ultra-intense lasers and ions or plasma from accelerators. The strong gravitational field is provided by huge accelerations of charges by ultra-intense lasers or accelerators. Finally, we conclude the perspective of astrophysics in strong EM fields and laboratory astrophysics.

\section{QED in EM Fields in the Minkowski Spacetime}\label{sec2}

In the Minkowski spacetime, the vacuum polarization dated back to Heisenberg and Euler, who calculated the one-loop effective action from the solution of electrons in a constant parallel electric and magnetic fields and then used the gauge invariance of the Maxwell theory to find the one-loop effective action.\cite{Heisenberg:1935qt} The effective action for spinless charged scalars was obtained by Weisskopf.\cite{Weisskopf:1996bu}
The general formulation of the one-loop action in a constant EM field was derived by Schwinger both for spinor and scalar QED, who introduced the proper-time integral representation.\cite{Schwinger:1951nm} The quantum field theory underlying the Heisenberg-Euler and Schwinger effective action may be found in Ref.~\citenum{Dittrich:1985yb}, physics in strong EM fields in Ref.~\citenum{Greiner:1985ce}, and the worldline formalism for QED action in Ref.~\citenum{Schubert:2001}.

In curved spacetimes, however, the task of calculating the one-loop effective actions even in a constant EM field becomes nontrivial and challenging. The vast literature focuses on perturbative expansion of one-loop effective action (for review and references, see Ref.~\citenum{Birrell:1982ix}). The graviton contributions to the Maxwell theory were found perturbatively.\cite{Dalvit:2000ay}
The worldline formalism was employed to find the effective actions up to quartic order of the Maxwell scalar and pseudo-scalar in general curved spacetimes.\cite{Bastianelli:2008cu,Davila:2009vt,Avramidi:2009wk} The exact one-loop effective action in a uniform electric field was first calculated in the $\mathrm{AdS}_2$ space in Ref. \citenum{Pioline:2005pf} and in the $\mathrm{(A)dS}_2$ in Ref.~\citenum{Cai:2014qba}. The QED action reduces to the Heisenberg-Euler and Schwinger action in the Minkowski spacetime when the spacetime curvature vanishes.

Considering the vantage and perspective of the method employed in Ref.~\citenum{Cai:2014qba}, it is worth introducing the in-out formalism from the Schwinger's variation principle, in which the one-loop effective action is given by the scattering amplitude $e^{i \int \sqrt{-g}d^4x {\cal L}_{\rm eff}} = \langle {\rm out} \vert {\rm in} \rangle$. Assuming that the in- and the out-vacua are well-defined in the asymptotic regions and the Bogliubov transformation between them is found explicitly, the exact one-loop effective action is formally given by (in natural unit of $c = \hbar =1$, and  $\alpha = e^2/4 \pi$)\cite{DeWitt:1975ys,DeWitt:2003pm}
\begin{eqnarray}
{\cal W}_{\rm eff} = \int \sqrt{-g}d^4x {\cal L}_{\rm eff} = \pm i {\cal V} T\sum_{\bf K} \ln (\alpha^*_{\bf K}), \label{formal action}
\end{eqnarray}
where ${\bf K}$ denotes all quantum numbers for the field, the plus (minus) sign is for scalar (spinor) field, and ${\cal V}$ and $\mathrm{T}$ are the relevant volume and time for the interaction.
The formal expression (\ref{formal action}) for one-loop action does not provide us with useful information about the vacuum polarization (real part of the effective action). When pairs are spontaneously produced from EM fields and/or curved spacetimes, Eq.~(\ref{formal action}) gives the vacuum persistence amplitude (twice the imaginary part)
\begin{eqnarray}
2 \Im \bigl({\cal W}_{\rm eff} \bigr) = \pm {\cal V} T \sum_{\bf K} \ln (1 \pm {\cal N}_{\bf K}), \label{vacuum persistence}
\end{eqnarray}
where ${\cal N}_{\bf K} = \vert \beta_{\bf K} \vert^2$ is the mean number of produced pairs and the upper (lower) sign is for scalars (fermions).

Noting that the Bogoliubov coefficients are given by simple ratios of gamma functions for almost all exactly solved models in EM fields, such as a constant or Sauter-type electric or magnetic fields either in the Coulomb and vector potentials, the gamma-function regularization was introduced, which expresses Eq.~(\ref{formal action}) in terms of the proper-time integral.\cite{Kim:2008yt,Kim:2009pg,Kim:2011cx} As an illustration, let us consider a constant electric field parallel to another constant magnetic field. One can find a Lorentz frame where two fields are parallel to each other when the pseudo-scalar ${\cal G} = {\bf E} \cdot {\bf B} \neq 0$. Then, the spin-averaged Bogoliubov coefficients for spin-1/2 fermion with mass $m$ and charge $q$ in constant parallel EM fields are given by\cite{Kim:2012vr}
\begin{eqnarray}
\alpha^{(\sigma)}_n = \bigl(\alpha^{(\sigma)}_{(1)n} \alpha^{(\sigma)}_{(-1)n} \bigr)^{1/2}, \quad \beta^{(\sigma)}_n = \bigl(\beta^{(\sigma)}_{(1)n} \beta^{(\sigma)}_{(-1)n} \bigr)^{1/2},
\end{eqnarray}
where
\begin{eqnarray}
\alpha^{(\sigma)}_{(r)n} = e^{-i(p^*+p) \frac{\pi}{2}} \frac{\sqrt{2 \pi} e^{-i p^* \frac{\pi}{2}}}{\Gamma (-p)}, \quad \beta^{(\sigma)}_{(r)n} = - e^{-i p^* \pi}.
\end{eqnarray}
Here, $\sigma = \pm 1/2$ denotes the spin component along the magnetic field and $r = \pm 1$ denotes the spin-tensor component along the electric field, and
\begin{eqnarray}
p = - \frac{1+r}{2} + i \frac{m^2+ qB (2n+1 - 2 \sigma)}{2qE}.
\end{eqnarray}
Using the integral representation of $\ln (\Gamma (-p^*))$, applying the Cauchy residue theorem, and renormalizing the vacuum energy and charge, one obtains the one-loop effective action\cite{Kim:2012vr}
\begin{eqnarray}
{\cal L}_{\rm eff}^{\rm sp} &=& - i \Bigl(\frac{qB}{2 \pi} \Bigr) \Bigl(\frac{qE}{2 \pi} \Bigr) \sum_{n, \sigma, r} \int_{0}^{\infty} \frac{ds}{s} \Bigl[ \frac{e^{p^*s}}{1 - e^{-s}} - \cdots \Bigr] \nonumber\\
&=& - \frac{1}{2 (2\pi)^2}  \int_{0}^{\infty} \frac{ds}{s^3} e^{-m^2 s} \Bigl[\frac{qBs}{\tanh (qBs)} \frac{qEs}{\tan (qEs)} - 1 - \frac{(qs)^2}{3} \bigl(E^2 - B^2 \bigr) \Bigr]. \label{qed E-B}
\end{eqnarray}
In Eq.~(\ref{qed E-B}), the Cauchy  theorem is applied to a quarter circle in the first quadrant, which gives the vacuum polarization with a convergent integral. And, from the Cauchy residue theorem follows the vacuum persistence amplitude (\ref{vacuum persistence}) due to spontaneous production of electron-positron pairs
\begin{eqnarray}
2 \Im \bigl({\cal L}_{\rm eff} \bigr) = \Bigl(\frac{qB}{2 \pi} \Bigr) \Bigl(\frac{qE}{2 \pi} \Bigr) \sum_{k = 1}^{\infty} \frac{1}{k} e^{- \frac{\pi m^2}{qE}k} {\rm cotanh} \Bigl( \frac{\pi E}{B} k \Bigr). \label{vac E-B}
\end{eqnarray}
The exact one-loop effective action (\ref{qed E-B}) is the proper-time integral of the Schwinger QED action. One can express the vacuum polarization (\ref{qed E-B}) in terms of gauge invariant scalar ${\cal F}$ and pseudo-scalar ${\cal G}$
\begin{eqnarray}
{\cal F} = \frac{1}{4} F_{\mu \nu} F^{\mu \nu} = \frac{1}{2} \bigl({\bf E}^2 - {\bf B}^2 \bigr), \quad {\cal G} = \frac{1}{4} F_{\mu \nu} {}^*F^{\mu \nu} = {\bf E} \cdot {\bf B},
\end{eqnarray}
and in a Lorentz frame where ${\bf E} \parallel {\bf B}$
\begin{eqnarray}
X \equiv \sqrt{2 ({\cal F} + i {\cal G})} = \vert {\bf B} + i {\bf E} \vert =  B+ i E.
\end{eqnarray}
Then, in any Lorentz frame, the electric and magnetic fields are given by\cite{Ruffini:2009hg}
\begin{eqnarray}
B = \Re[X] = \sqrt{({\cal F}^2 + {\cal G}^2 )^{1/2} + {\cal F}}, \quad  E = \Im [X] =  \sqrt{({\cal F}^2 + {\cal G}^2 )^{1/2} - {\cal F}}.
\end{eqnarray}
and the effective action of the Maxwell theory and the one-loop QED action takes the Lorentz-invariant form\cite{Schwinger:1951nm}
\begin{eqnarray}
{\cal L}_{\rm eff} &=& \frac{1}{4 \pi} {\cal F} - \frac{1}{2 (2\pi)^2}  \int_{0}^{\infty} \frac{ds}{s^3} e^{-m^2 s} \Bigl[(qs)^2 {\cal G} \frac{\Re[\cosh(qXs) ]}{\Im[\cosh(qXs)]} - 1 - \frac{2 (qs)^2}{3} {\cal F} \Bigr] \nonumber\\
&=& \frac{1}{8 \pi} \bigl({\bf E}^2 - {\bf B}^2 \bigr) + \frac{1}{90 (2 \pi)^2} \Bigl(\frac{q}{m} \Bigr)^2
\bigl[ \bigl({\bf E}^2 - {\bf B}^2 \bigr) + 7 \bigl( {\bf E} \cdot {\bf B} \bigr)^2 \bigr] + \cdots. \label{qed gen}
\end{eqnarray}

In a supercritical magnetic field $(\beta = m^2/2qB \leq 1)$, the QED action can be expressed in the Hurwitz zeta function
\begin{eqnarray}
{\cal L}_{\rm eff}^{\rm sp} (\beta) =  \Bigl( \frac{m^2}{4 \pi \beta} \Bigr)^2 \Bigl[ 2 \zeta'(-1, \beta) -
(\beta^2 - \beta + \frac{1}{6}) \ln (\beta) + \frac{\beta^2}{2} - \frac{1}{6} \Bigr],
\label{zeta B-act}
\end{eqnarray}
which was transformed of the integral (\ref{qed E-B}) in Ref.~\citenum{Dittrich:1975au} and directly obtained from the gamma function in Ref.~ \citenum{Kim:2014iia}. The QED action in a supercritical electric field $(\epsilon = m^2/2qE \leq 1)$ is also expressed as\cite{Kim:2014iia}
\begin{eqnarray}
\Re[{\cal L}_{\rm eff}^{\rm sp} (\epsilon)] &=&  - \Bigl( \frac{m^2}{4 \pi \epsilon} \Bigr)^2 \Bigl[\zeta'(-1, i \epsilon) + \zeta'(-1, - i \epsilon)
+ (\epsilon^2 - \frac{1}{6}) \ln (\epsilon) - \frac{\pi \epsilon}{2} - \frac{\epsilon^2}{2} - \frac{1}{6} \Bigr], \nonumber\\
\Im[{\cal L}_{\rm eff}^{\rm sp} (\epsilon)] &=&  \Bigl( \frac{m^2}{4 \pi \epsilon} \Bigr)^2 \Bigl[i \zeta'(-1, i \epsilon) -i \zeta'(-1, - i \epsilon) + \frac{\pi}{2} \Bigl(\frac{1}{6}- \epsilon^2 \Bigr) - \epsilon \ln (\epsilon) \Bigr].
\label{zeta E-act}
\end{eqnarray}
The QED action (\ref{zeta B-act}) shows a power-law behavior of $\beta$ for supercritical fields. The vacuum persistence amplitude in Eq.~(\ref{zeta E-act}) recovers Eq.~(\ref{vac E-B}) in the limit of $B = 0$.  Remarkably, there is a numerical threshold around $E_\mathrm{th} \approx (e^{-1/10}/2) E_c$ for QED action in the electric field, which may be a consequence of pair production but require further study.

A few comments are in order. In applying the in-out formalism to strong EM fields, a proper boundary condition should be imposed on quantum fields depending on the choice of gauge potentials. A constant electric field has either a Coulomb potential or a vector potential. In the former case the quantum field becomes the tunneling problem under an inverted harmonic barrier, to which the boundary condition from the causality (group velocity) applies.\cite{Nikishov:1970,Kim:2003qp} Then, for the production of scalars the relative ratio of the incident flux to the reflected flux gives the vacuum persistence amplitude while the ratio of the transmitted flux to the reflected flux gives the mean number (an amplification factor) for pairs. For the production of fermions, the relative ratio of the reflected flux to the incident flux leads to the vacuum persistence amplitude and the ratio of transmitted flux to the incident flux to the mean number of pairs. On the other hand, in the latter case the quantum field becomes the scattering problem over a barrier, and an incoming positive frequency solution is scattered by the barrier to split into a branch of positive frequency solution and another branch of negative frequency solution. The Schwinger pair production is similar to particle production in an expanding universe.\cite{Parker:1969au} The final expression of the vacuum polarization and Schwinger effect is independent of the gauge choices as expected. For QED problem in curved spacetimes, the choice of vector potential for a uniform electric field and the scattering boundary condition is convenient for a charged quantum field in dS space, as will be shown in Sec.~\ref{sec3}. The axisymmetric KN black holes or the static RN black holes have the tunneling boundary condition for charged fields in Sec.~\ref{sec4}.

Produced pairs backreact to the background fields when the Schwinger effect becomes dominant. In the Minkowski spacetime, this happens when an electric field is close to the critical strength; one pair is created per unit Compton volume and per unit Compton time of the particle. The energy density of produced pairs equals to that of the electric field itself, and the energy conservation requires the backreaction (induced four-current) of produced pairs to be part of the governing equation. Simply stated, positive and negative charges accelerating by the field induce a current increasing in time, which generates a magnetic field increasing in time. The induced magnetic field in turn induces an electric field in the opposite direction to the field, and thus reduces the combined total electric fields. It is a consequence of the energy conservation because the produced pairs carry part of the energy of background field. In fact, there is an overshooting problem of produced pairs leading to a plasma oscillation, and the pair annihilation into photons thermalizes plasma of pairs and photons as a final state as shown in Refs.~\citenum{Kluger:1991ib} and \citenum{Ruffini:2003} (for a comprehensive review and references, see Ref.~\citenum{Ruffini:2009hg}).

\section{QED in $\mathrm{(A)dS}$ Space and the Effective Temperature}\label{sec3}

A generic expanding universe, such as the inflationary universe driven by an inflaton or the $\mathrm{dS}$ space by the vacuum energy, produces particles.\cite{Parker:1969au} A quantum field in $\mathrm{dS}$ spaces has a thermal state of the Gibbons-Hawking temperature $T_\mathrm{GH} = H/2 \pi$ (in geometrized unit of $G = k_\mathrm{B} = 1$).\cite{Gibbons:1977mu} Thus, the $\mathrm{dS}$ space emits all species of particles including charged pairs through Hawking radiation from the cosmological horizon, which provides a cosmic laser.\cite{Polyakov:2009nq} Hence, the Schwinger effect and vacuum polarization in $\mathrm{dS}$ space is not only theoretically interesting but also has cosmological implications because of cosmic scale magnetic fields and phase transitions of particle physics models in the early universe.

On the other hand, the $\mathrm{AdS}$ space is another maximally symmetric solution of the Einstein equation with a negative vacuum energy. The $\mathrm{AdS}$ black holes have the $\mathrm{AdS}$ space as an asymptotic boundary and a positive heat capacity in contrast to the black hole with an asymptotically flat space.\cite{Hawking:1983} Though the accelerating phases of the universe is close to the $\mathrm{dS}$ space, the $\mathrm{AdS}$ space is the near-horizon geometry of (near-)extremal black holes. The near-horizon geometry is locally a warped product of $\mathrm{AdS}_2 \times \mathrm{S}^2$ for a near-extremal RN black hole and another warped product of $\mathrm{AdS}_3$ for a near-extremal KN black hole,\cite{Kunduri:2007vf,Kunduri:2008rs,Kunduri:2013ana} as shown in Table~\ref{aba:tbl1} and will be shown in detail in Sec.~\ref{sec4}.

\begin{table}
\tbl{QED in $\mathrm{(A)dS}_2$ from black holes}
{\begin{tabular}{@{}ccc@{}}
\toprule
Black holes  &  Parameter of black holes & Near-horizon geometry  \\
\hline\\[-6pt]
RN black holes & Near-extremal black hole  & $\mathrm{AdS}_2 \times S^2$ [\citenum{Chen:2012zn,Chen:2014yfa}] \\
 & nonextremal black hole & $\mathrm{Rindler}_2 \times S^2$ \\
\hline\\[-6pt]
Rotating black hole in $\mathrm{dS}$ & $\mathrm{S}$-scalar wave & QED in $\mathrm{dS}_2$ [\citenum{Anninos:2009jt,Bredberg:2009pv}] \\
\botrule
\end{tabular} \label{aba:tbl1}
}
\end{table}

The symmetry of $\mathrm{(A)dS}$ space allows analytical solutions of a charged scalar or fermion field in a uniform electric field. The $\mathrm{(A)dS}$ space with a uniform electric field is not a solution of the Einstein-Maxwell theory, so we assume that the energy density of the electric field is smaller than the vacuum energy density, $E \ll H (K)$ with $H = \sqrt{\Lambda/3}$ and thereby the backreaction of produced pairs through the Schwinger mechanism is not sufficient to modify the $\mathrm{(A)dS}$ space. Under this assumption, the Schwinger effect was studied in the $\mathrm{dS}_2$ space,\cite{Garriga:1994bm,Kim:2008xv} and in the $\mathrm{AdS}_2$ space.\cite{Pioline:2005pf,Kim:2008xv} In the presence of an electric field, the Schwinger effect has an analytical continuation between the scalar curvature $R_{\mathrm dS} = 2 H^2$ and $R_{\mathrm AdS} = - 2 K^2$.\cite{Kim:2008xv} The vacuum polarization was also found in the $\mathrm{AdS}_2$ space\cite{Pioline:2005pf} and in the $\mathrm{(A)dS}_2$ space.\cite{Cai:2014qba}

The vacuum polarization and the Schwinger effect in the $\mathrm{dS}_2$ space, as exploited in Sec.~\ref{sec2}, was found in Ref.~\citenum{Cai:2014qba}. Further, it was proposed that the effective temperature
\begin{eqnarray}
T_\mathrm{eff} = T_\mathrm{U} + \sqrt{T^2_\mathrm{U} + T^2_\mathrm{GH}}, \label{eff tem}
\end{eqnarray}
with the Unruh temperature for an accelerating charge and an effective mass in the $\mathrm{dS}_2$ space
\begin{eqnarray}
 T_\mathrm{GH} = \frac{H}{2 \pi}, \quad  T_\mathrm{U} = \frac{qE/{\bar m}}{2 \pi}, \quad {\bar m} = m \sqrt{1 - \Bigl(\frac{H}{2m} \Bigr)^2},
\end{eqnarray}
explains the leading Boltzmann factor $e^{-{\bar m}/T_\mathrm{eff}}$ for produced pairs. This is an analog of the effective temperature for an accelerating observer in the $\mathrm{(A)dS}_2$ space\cite{Narnhofer:1996zk,Deser:1997ri}
\begin{eqnarray}
T_\mathrm{eff} = \sqrt{T^2_\mathrm{U}+ \frac{R_\mathrm{(A)dS}}{2 (2 \pi)^2} }. \label{Unruh}
\end{eqnarray}
The Breitenlohner-Freedman (BF) bound for the stability of the $\mathrm{AdS}_2$ space\cite{Breitenlohner:1982jf} becomes  $T_\mathrm{U}^2 < -R_\mathrm{AdS}/8 \pi^2$ in the presence of a uniform electric field, so the spontaneous pair production violates the BF bound.\cite{Pioline:2005pf}
The additional $T_\mathrm{U}$ in Eq.~(\ref{eff tem}) is due to a ultra-relativistic effect of the Schwinger mechanism. The effective temperature is consistent with the Schwinger effect in the $\mathrm{(A)dS}_2$ space,\cite{Kim:2008xv} and has a direct interpretation of the charge emission from near-extremal charged black holes, as will be shown in Sec.~\ref{sec4}.

The Schwinger pair production of scalars has been studied in the $\mathrm{dS}_4$  space in Ref.~\citenum{Kobayashi:2014zza} and of fermions in Refs.~\citenum{Stahl:2015cra,Hayashinaka:2016qqn}, and of scalars in the $\mathrm{dS}_2$ space in Refs.~\citenum{Frob:2014zka,Banyeres:2018aax,Rajeev:2019okd}. Here, we follow Ref.~\citenum{Kim:2016dmm}, in which the Schwinger mechanism for bosons and fermions was discussed in any dimensional dS space. But we shall confine our study to the Schwinger effect and vacuum polarization of scalars and fermions in the $\mathrm{(A)dS}_4$ space and discuss the analytic continuation between the $\mathrm{dS}_4$ and $\mathrm{AdS}_4$ spaces. Expressing the density of produced pairs per unit four-volume in terms of the scalar curvature $R_4 = 12 H^2$, Eq.~(16) of Ref.~\citenum{Kim:2016dmm} reads
\begin{eqnarray}
\frac{d^4{\cal N}}{dx^4} = {\cal D} (E) \Bigl( \frac{e^{- {\cal S}_{\mu} + {\cal S}_{\lambda}} \pm e^{- 2 {\cal S}_{\mu} } }{1 - e^{- 2{\cal S}_{\mu}}} \Bigr), \quad {\cal D} (E) = \frac{(2 |\sigma| +1) R_4 {\cal S}_{\mu}}{24 (2 \pi)^2} \int \frac{d^2 {\bf k}_{\perp}}{(2 \pi)^2}, \label{sch ds}
\end{eqnarray}
where ${\cal D}$ is the density of states and the relativistic instanton actions are
\begin{eqnarray}
{\cal S}_{\mu} = 2 \pi \Bigl( \frac{12qE}{R_4} \Bigr) \sqrt{1+ \frac{{\bar m}^2 R_4}{12 (qE)^2}}, \quad {\cal S}_{\lambda} = 2 \pi \Bigl( \frac{12qE}{R_4} \Bigr) \frac{1}{ \sqrt{1+ \frac{R_4}{12 (qE)^2} {\bf k}^2_{\perp}}}. \label{ins act}
\end{eqnarray}
Here, the upper (lower) sign is for scalars (fermions) and ${\bar m} = m \sqrt{1 - 3R_4/16m^2}$ is an effective mass in the $\mathrm{(A)dS}_4$ space. In arriving at Eq.~(\ref{sch ds}) the integral along the longitudinal momentum is performed by the saddle point approximation. Note that
the Schwinger formulae (\ref{sch ds}) respect the translational symmetry along the electric field direction as well as the transverse directions.
The leading Boltzmann factor $e^{- {\cal S}_{\mu} + {\cal S}_{\lambda}}$ allows an analytic continuation from the $\mathrm{dS}_4$ space to the $\mathrm{AdS}_4$ space, in which $R_4$ changes the signs of ${\cal S}_{\mu}$ and ${\cal S}_{\lambda}$. The other terms in Eq.~(\ref{sch ds}) should have $|{\cal S}_{\mu}|$ for the $\mathrm{AdS}_4$ space.

A few comments are in order. The Schwinger formulae (\ref{sch ds}) for scalars and fermions reduce to those in the Minkowski spacetime when $R_4 = 0$. On the other hand, in the zero-field limit $(E= 0)$, Eq.~(\ref{sch ds}) recovers Hawking radiation with the Gibbons-Hawking temperature.\cite{Kim:2016dmm} Integrating the transverse momenta of Eq.~(\ref{sch ds}) up through quadratic order, we obtain approximately the Schwinger formula
\begin{eqnarray}
\frac{d^4{\cal N}}{dx^4} \simeq \frac{2 |\sigma| +1}{4 \pi} \Bigl(\frac{qE}{2 \pi} \Bigr)^2 e^{- \frac{{\bar m}}{T_\mathrm{eff}}},
\end{eqnarray}
where $T_\mathrm{eff}$ is the effective temperature (\ref{eff tem}). The reason for the presence of Eq.~(\ref{eff tem}) in the $\mathrm{(A)dS}_2$ space is that the motion of charges occurs essentially along the direction of the electric field regardless of the dimensions of spacetimes, which will also be shown for near-extremal black holes in Sec.~\ref{sec4}. Provided that the electric field is strong enough to rapidly accelerate charged pairs to the speed of light but still smaller than the vacuum energy $E \ll H (K)$ so that the energy density of the field does not modify the dS space, the induced current density per unit four-volume is approximately given by $J = 2q (d^4{\cal N}/dx^4)$.

The scalar QED action consistent with Eq.~(\ref{vacuum persistence}) is given by
\begin{eqnarray}
{\cal L}_\mathrm{eff}^\mathrm{sc} = {\cal D} (E) \int_0^{\infty} \frac{ds}{s} \Bigl[ e^{- ({\cal S}_{\mu} - {\cal S}_{\lambda})s/2 \pi} \Bigl( \frac{1}{\sin(s/2)} - \frac{2}{s} \Bigr) -  e^{- {\cal S}_{\mu} s/\pi } \Bigl( \frac{\cos(s/2)}{\sin(s/2)} - \frac{2}{s} \Bigr)  \Bigr],\nonumber\\ \label{sc act}
\end{eqnarray}
and the spinor QED action by
\begin{eqnarray}
{\cal L}_\mathrm{eff}^\mathrm{sp} = - {\cal D} (E) \int_0^{\infty} \frac{ds}{s} e^{- {\cal S}_{\mu} s/2 \pi } \bigl( e^{{\cal S}_{\lambda}s/2 \pi} - e^{- {\cal S}_{\mu} s/2 \pi }  \bigr)
 \Bigl( \frac{\cos(s/2)}{\sin(s/2)} - \frac{2}{s} \Bigr). \label{sp act}
\end{eqnarray}
Note that when $R_4 =0$, the QED actions (\ref{sc act}) and (\ref{sp act}) reduce to the Heisenberg-Euler-Schwinger actions
\begin{eqnarray}
{\cal L}_\mathrm{eff} &=& \pm \frac{[2] qE}{2(2 \pi)^2} \int \frac{d^2 {\bf k}_{\perp}}{(2 \pi)^2} \int_0^{\infty} \frac{ds}{s} e^{- \pi \frac{m^2 + {\bf k}_{\perp}^2}{qE}s } \Bigl( \frac{[\cos(s/2)]}{\sin(s/2)} - \frac{2}{s} \mp \frac{[2]s}{12} \Bigr), \label{HES act}
\end{eqnarray}
where the upper (lower) sign is for scalar (spinor) QED and the square brackets become unity for scalars. More interesting results can be obtained by taking the zero-field limit $(E = 0)$, which is the one-loop effective action for the $\mathrm{dS}_4$ space
\begin{eqnarray}
{\cal L}_\mathrm{eff} = \pm \frac{(2 |\sigma|+1) 6H^3 {\bar m}}{(2 \pi)^2} \int_0^{\infty} \frac{ds}{s} e^{- {\cal S}_{\mu} s/2 \pi } \Bigl( \frac{[\cos(s/2)]}{\sin(s/2)} - \frac{2}{s} \Bigr). \label{ds act}
\end{eqnarray}
The reversed role of the spectral functions for scalars and fermions in pure dS space compared with those in Minkowski spacetime is a consequence of the difference between the Bose-Einstein or Fermi-Dirac distribution and the Boltzmann distribution in the vacuum persistence amplitude (\ref{vacuum persistence}).\cite{Stephens:1989fb} In the zero-field limit, the integrals are independent of the transverse momenta and the transverse momenta cut-off $k_\mathrm{c} = 2 \pi/H$ in Eqs.~(\ref{sc act}) and (\ref{sp act}) is given by the Hubble constant, the inverse of the Hubble radius, because the quantum states for charges are defined when their wavelengths fall within the Hubble horizon. The subtracted term corresponds to renormalization of the vacuum energy. The series expansion of the parenthesis in Eq.~(\ref{ds act}) gives the effective action in all the powers of $R_4$, which start with $(R_4)^2$ and drastically differ from the perturbative result, for instance, in Ref.~\citenum{Birrell:1982ix}. The power series expansion of the exact one-loop effective action is admissible because any derivative of higher curvatures can always be expressed in terms of the Hubble constant. It will be interesting to compare the action (\ref{ds act}) in planar coordinates with Eq.~(3.12) in global coordinates of Ref.~\citenum{Das:2006wg}.

Finally, we discuss the effect of a magnetic field on the Schwinger effect studied in Ref.~\citenum{Bavarsad:2017oyv}. In contrast to the Minkoswki spacetime, any Lorentz boost changes the conformal factor of the $\mathrm{dS}$ space as well as the conformal time, so there is no comoving observer who can see a magnetic field parallel to an electric field for a given configuration of EM field. Assuming a uniform magnetic field parallel to another uniform electric field to a comoving observer and performing the saddle point approximation of the momentum along the direction of fields, the Schwinger effect is given by the density of pairs per unit four-volume in the comoving coordinates
\begin{eqnarray}
\frac{d^4 {\cal N}}{dx^4}= {\cal D} \sum_{n = 0}^{\infty}
\Bigl[ \frac{e^{{\cal S}_{\lambda}} -1}{e^{{\cal S}_{\lambda}} - e^{- {\cal S}_{\mu}}} + \frac{1}{ e^{{\cal S}_{\mu}} - 1 }\Bigr], \quad {\cal D} (E,B) = \Bigl(\frac{qB}{ (2 \pi) \Omega^2} \Bigr) \Bigl(\frac{R_4 {\cal S}_{\mu}}{12 (2 \pi)^2} \Bigr), \label{B sch}
\end{eqnarray}
where $\Omega$ is the conformal factor for the comoving observer, $B$ denotes a conserved flux $B_0 \Omega^2 = B$ in the cosmological spacetime, ${\cal S}_{\mu}$ is in Eq.~(\ref{ins act}), and
\begin{eqnarray}
{\cal S}_{\lambda} = 2 \pi \Bigl( \frac{12 qE}{R_4} \Bigr) \sqrt{\frac{\frac{12(qE)^2}{R_4}}{\frac{12(qE)^2}{R_4} + (2n+1)qB}}. \label{ll act}
\end{eqnarray}
The transverse momenta in the instanton action (\ref{ins act}) is replaced by Landau levels as expected from the motion of charges in a constant magnetic field. The prefactor ${\cal D}$ is the density of states and the summation is the mean number of produced pairs in the Landau levels. In the limit $R_4 = 0$, Eq.~({\ref{B sch}) reduces to the Schwinger formula in the Minkowski spacetime, and a proper summation over Landau levels and taking the $B=0$ limit leads to the Schwinger effect in a pure electric field. The zeta-function regularization yields
\begin{eqnarray}
\frac{d^4 {\cal N}}{dx^4}= {\cal D} (E,B)
\Bigl(\frac{1}{e^{2 {\cal S}_{\mu}} - 1} \Bigr) \Bigl[ \frac{1}{2}  + \sum_{n = 0}^{\infty} e^{{\cal S}_{\mu} + {\cal S}_{\lambda}} \Bigr]. \label{B sch2}
\end{eqnarray}
The QED action\cite{Bavarsad:2019ab} consistent with the pair production rate (\ref{B  sch}) can be found from the Bogoliubov coefficient according to Sec.~\ref{sec2}.

A few comments are in order again. In the zero-field limit $(E=B=0)$ the formula (\ref{B sch2}) recovers Hawking radiation with the Gibbons-Hawking temperature. It was argued that the induced current from the Schwinger effect has a surprising phenomenon, the so-called IR-hyperconductivity, in which the current increases even though the electric field decreases.\cite{Frob:2014zka} In the weak magnetic field limit, the regularized induced current is given by (recovering $H$),\cite{Bavarsad:2017oyv}
\begin{eqnarray}
J_\mathrm{reg} = \frac{q}{H} \Bigl(\frac{qB}{ 2 \pi \Omega^2} \Bigr) \Bigl(\frac{qE}{ 2 \pi} \Bigr) \frac{\sinh({\cal S}_{\lambda})}{\sin({\cal S}_{\mu})}.
\end{eqnarray}
In the infrared regime, the induced current is approximated by
\begin{eqnarray}
J_\mathrm{reg} = \frac{9 q}{2} \Bigl(\frac{qB}{ 2 \pi \Omega^2} \Bigr) \Bigl(\frac{qE}{ 2 \pi} \Bigr) \frac{H^3}{(qE)^2 + ({\bar m}H)^2},
\end{eqnarray}
and increases even if the electric field decreases until $qE = {\bar m}H$ and then reaches the maximum
\begin{eqnarray}
J_\mathrm{reg} = \frac{9 qH^2}{4 (2 \pi) {\bar m}} \Bigl(\frac{qB}{ 2 \pi \Omega^2} \Bigr).
\end{eqnarray}
A simple interpretation is that Hawking radiation resides and enhances the induced current regardless of the strength of the electric field.

\section{QED in Charged Black Holes}\label{sec4}

The general axisymmetric solution of the Einstein equation in four dimensions is the KN black hole which has four parameters: mass $M$, angular momentum per unit mass $a = J/M$, electric charge $Q$, and magnetic charge $M$.\cite{Stephens:1989fb} The KN metric in the Boyer-Lindquist coordinates is given by
\begin{eqnarray}
ds^2 = - \frac{\Delta}{\rho^2} \bigl(dt - a \sin^2 \theta d \phi \bigr)^2 + \frac{\rho^2}{\Delta} dr^2 + \rho^2 d \theta^2 + \frac{\sin^2 \theta}{\rho^2} \bigl[ (r^2 + a^2) d \phi - a dt \bigr]^2, \label{KN QP}
\end{eqnarray}
where
\begin{eqnarray}
\rho^2 = r^2 + a^2 \cos^2 \theta, \quad \Delta = \bigl(r - M)^2 + \bigl(a^2 + Q^2 + P^2 - M^2 \bigr).
\end{eqnarray}
The gauge field for charges takes the one-form
\begin{eqnarray}
{\bf A}_{[1]} = - \Bigl(\frac{Qr - P a \cos \theta}{\rho^2} \Bigr)dt + \Bigl(\frac{Qa r \sin^2 \theta + P\bigl(\pm \rho^2 - \cos \theta (r^2 +a^2) \bigr) }{\rho^2} \Bigr) d \phi. \label{gauge}
\end{eqnarray}
The Hodge dual field has another one-form
\begin{eqnarray}
\overline{\bf A}_{[1]} = - \Bigl(\frac{Pr + Q a \cos \theta}{\rho^2} \Bigr)dt + \Bigl(\frac{Pa r \sin^2 \theta - Q\bigl(\pm \rho^2 - \cos \theta (r^2 +a^2) \bigr) }{\rho^2} \Bigr) d \phi,
\end{eqnarray}
which corresponds to $Q \rightarrow P$ and $P \rightarrow -Q$ of ${\bf A}_{[1]}$.
A charged scalar field $\Phi$ with the mass $m$, electric charge $q$, and magnetic charge $p$ can be separated  as
\begin{eqnarray}
\Phi = e^{- i \omega t + i (n \pm (q P- pQ))\phi} R(r) S(\theta),
\end{eqnarray}
where $R(r)$ satisfies the radial equation\cite{Semiz:1991kh}
\begin{eqnarray}
\frac{d}{dr} \Bigl(\Delta \frac{dR}{dr}\Bigr) + \Bigl[\frac{\bigl( (r^2 + a^2) \omega - (q Q+ pP) r - a n \bigr)^2}{\Delta} -m^2 (r^2+ a^2) + 2 an \omega - \lambda   \Bigr] R = 0, \nonumber\\ \label{sep eq}
\end{eqnarray}
and $S(\theta)$ is the monopole spheroidal harmonics with the eigenvalue $\lambda$ and $\omega$ is the energy. The black hole solution has the limits as shown in Table~\ref{aba:tbl2}: $a = 0$ corresponds to nonrotating charged RN black holes with $Q$ and $P$, which in turn has a purely electrically charged black hole $Q \neq 0$ and $P=0$ or magnetically charged one $P \neq 0$ and $Q=0$. The separation constant for RN black holes is $\lambda = l(l+1)$ for spherical harmonics. The eigenvalues $\lambda$ for KN black holes can be found numerically.

\begin{table}
\tbl{Black holes with $Q$ and $P$ and the emission of $p$ and $q$.}
{\begin{tabular}{@{}cccccc@{}}
\toprule
Black holes &  $a = \frac{J}{M}$ & Q & P & Hawking radiation & Schwinger effect \\
\hline\\[-6pt]
 & a $\neq$ 0 & Q $\neq$ 0 & P $\neq$ 0 & dyons q \& p & dyons q \& p \\
KN & a $\neq$ 0 & Q $\neq$ 0 & P = 0 & charges q [\citenum{Hawking:1974sw}] & charges q  \\
 & a $\neq$ 0 & Q = 0 & P $\neq$ 0 & monopoles p & monopoles p \\
\hline\\[-6pt]
Near-extremal KN & a $\neq$ 0 & Q $\neq$ 0 & P $\neq$ 0 & charges q \& p & charges q \& p [\citenum{Chen:2017mnm,Chen:2019rtd}]  \\
 & a $\neq$ 0 & Q $\neq$ 0 & P = 0 & charges q [\citenum{Page:1976}] & charges q  [\citenum{Chen:2016caa}]  \\
\hline\\[-6pt]
 & a = 0 & Q $\neq$ 0 & P $\neq$ 0 & dyons q \& p & dyons q \& p \\
RN & a = 0 & Q $\neq$ 0 & P = 0 & charges q [\citenum{Hawking:1974sw}] & charges q  \\
 & a = 0 & Q = 0 & P $\neq$ 0 & monopoles p & monopoles p \\
 \hline\\[-6pt]
Near-extremal RN & a = 0 & Q $\neq$ 0 & P = 0 & charges q & charges q [\citenum{Chen:2012zn}]  \\
\botrule
\end{tabular} \label{aba:tbl2}
}
\end{table}

The KN black holes has the event horizon and the Cauchy horizon
\begin{eqnarray}
r_\mathrm{H} = M + \sqrt{M^2 - \bigl(a^2 + Q^2 + P^2 \bigr)}, \quad r_\mathrm{C} = M - \sqrt{M^2 - \bigl(a^2 + Q^2 + P^2 \bigr)}.
\end{eqnarray}
The thermodynamical variables for dyonic KN black holes are the Hawking temperature and the Bekenstein-Hawking entropy\cite{Chen:2017mnm}
\begin{equation}\label{TSKN}
T_\mathrm{H} = \frac{r_\mathrm{H} - r_\mathrm{C}}{4 \pi (r_\mathrm{H}^2 + a^2)}, \quad S_\mathrm{BH} = \pi (r_\mathrm{H}^2 + a^2),
\end{equation}
and the electric and magnetic chemical potentials, and the angular velocity
\begin{equation}
\Phi_\mathrm{H} = - \frac{Q r_\mathrm{H}}{r_\mathrm{H}^2 + a^2}, \quad {\bar\Phi}_\mathrm{H} = - \frac{P r_\mathrm{H}}{ r_\mathrm{H}^2 + a^2}, \quad \Omega_\mathrm{H} = \frac{a}{ r_\mathrm{H}^2 + a^2}.
\end{equation}
The emission of charges from charged black holes was independently studied as the Schwinger mechanism.\cite{Zaumen:1974,Carter:1974yx,Damour:1974qv,Gibbons:1975kk,Damour:1976jd} Since then, the Schwinger effect from a charged black hole has been revisited from the tunneling picture of virtual particles in the Dirac sea whose mass gap is modified by not only the electric potential but also the gravitational potential.\cite{Ternov:1986bf,Khriplovich:1998si,Khriplovich:1999gm,Khriplovich:2002qn,Kim:2004us,Ruffini:2013hia}
The leading Boltzmann factor can be simply derived by the phase-integral method\cite{Kim:2007pm} or complex path analysis,\cite{Srinivasan:1998ty} which searches for a solution of the form $R=e^{i {\cal S} (r)/[\hbar]}$ for the radial equation (\ref{sep eq}). Then, the WKB approximation
\begin{eqnarray}
{\cal S} = \int^r dr \frac{\bigl( (r^2 + a^2) \omega - (q Q+ pP) r - a n \bigr)}{(r - r_\mathrm{H})(r - r_\mathrm{C})}
\end{eqnarray}
has a residue contribution from the simple pole at $r_\mathrm{H}$ in the complex plane of $r$, and the decay of the field amplitude $|\Phi|^2$ is given by the leading Boltzmann factor of Hawking radiation for dyons
\begin{eqnarray}
e^{- 2 \Im {\cal S}} = \exp \Bigl[- \frac{\omega - a \Omega_\mathrm{H} - q \Phi_\mathrm{H} - p {\bar \Phi}_\mathrm{H}}{T_\mathrm{H}} \Bigr].
\end{eqnarray}
Note that only the pairs produced near $r_\mathrm{H}$ contribute to Hawking radiation because $r_\mathrm{C}$ is inside $r_\mathrm{H}$ farther away than the Compton wavelength of the particle and pairs produced near $r_\mathrm{C}$ cannot escape beyond $r_\mathrm{H}$.

On the other hand, the emission of charged particles from near-extremal black holes is better understood by using the near-horizon geometry whose symmetry leads the solutions in terms of confluent hypergeometric functions for charged RN black holes and hypergeometric functions for KN black holes. The idea behind this concept is that the near-horizon geometry is an $\mathrm{AdS}_2$ space for RN black holes and a warped $\mathrm{AdS}_3$ space for KN black holes, in which QED problem, such as the Schwinger effect and/or vacuum polarization, can be analytically solved, as shown in Refs.~\citenum{Cai:2014qba,Chen:2012zn,Chen:2014yfa,Chen:2017mnm} and \citenum{Chen:2016caa}, and as explained in Sec.~\ref{sec3} and summarized in this proceedings~\citenum{Chen:2019rtd}. The near-horizon geometry of static or stationary spacetime in any dimension can be found in Ref.~\citenum{Kunduri:2013ana}. This section is complementary to Ref.~\citenum{Chen:2019rtd}.
To make use of the symmetry of the near-horizon geometry, one magnifies the near-horizon region, for instance, elongates the radial coordinates and shortens the time coordinate of RN black holes. For near-extremal rotating black holes, one first corotates all coordinates with the horizon angular velocity $\Omega_\mathrm{H}$
\begin{eqnarray}
\varphi \rightarrow \varphi + \Omega_\mathrm{H} t,
\end{eqnarray}
and then elongates or shortens $r, t$ and $M$ by using an infinitesimally small parameter $\varepsilon \rightarrow 0$ and introducing another parameter $B$ for the deviation from the extremality
\begin{eqnarray} \label{ex-bh lim}
r \rightarrow r_\mathrm{H} + \varepsilon r, \quad t \rightarrow \frac{r_\mathrm{H}^2 + a^2}{\varepsilon} t, \quad M \rightarrow r_\mathrm{H} + \varepsilon^2 \frac{B^2}{2 r_\mathrm{H}}.
\end{eqnarray}
Then, the near-horizon geometry of the near-extremal dyonic KN black hole has the metric\cite{Chen:2017mnm}
\begin{eqnarray} \label{ex-bh met}
ds^2 = \Gamma(\theta) \Bigl[ -(r^2 - B^2) dt^2 + \frac{dr^2}{r^2 - B^2} + d\theta^2 \Bigr] + \gamma(\theta) (d\varphi +  2 \Omega_\mathrm{H} r_\mathrm{H} r dt)^2,
\end{eqnarray}
and the gauge potential has the one-form
\begin{eqnarray}
A_{[1]} = \frac{ 2 P r_\mathrm{H} a \cos\theta -Q (r_\mathrm{H}^2 - a^2 \cos^2\theta)}{\Gamma(\theta)} r dt - \frac{Q r_\mathrm{H} a \sin^2\theta - P l_{\rm AdS}^2 \cos\theta \pm P\Gamma(\theta)}{\Gamma(\theta)} d\varphi, \nonumber\\
\label{ex-bh gaug}
\end{eqnarray}
where $l_{\rm AdS} = \sqrt{r_{\rm H}^2 + a^2}$ is the radius of the ${\rm AdS}_2$ space and
\begin{eqnarray}
\Gamma(\theta) = r_\mathrm{H}^2 + a^2 \cos^2\theta, \quad \gamma(\theta) = \frac{l_{\rm AdS}^4 \sin^2\theta}{r^2_\mathrm{H} + a^2 \cos^2\theta}.
\end{eqnarray}
The Hodge dual field is obtained by substituting $Q \rightarrow P$ and $P \rightarrow -Q$ in Eq.~(\ref{ex-bh gaug}).
The warped AdS$_3$ geometry in Eq.~(\ref{ex-bh met}) allows the dual CFTs description with complex weights for dynonic KN black holes.

The scaled Hawking temperature and the entropy are\cite{Chen:2017mnm}
\begin{eqnarray}\label{ex-bh tem-en}
T_\mathrm{H} = \frac{B}{2 \pi}, \quad S_\mathrm{BH} = \pi (r_\mathrm{H}^2 + a^2 + 2 B r_\mathrm{H}),
\end{eqnarray}
and the chemical potentials are
\begin{eqnarray}\label{ex-bh chem}
\Phi_\mathrm{H} = \frac{Q (Q^2 + P^2) B}{r_\mathrm{H}^2 + a^2}, \quad \bar\Phi_\mathrm{H} = \frac{P (Q^2 + P^2) B}{r_\mathrm{H}^2 + a^2}, \quad \Omega_\mathrm{H} = - \frac{2 a r_\mathrm{H} B}{r_\mathrm{H}^2 + a^2}.
\end{eqnarray}
The electrically and/or magnetically charged RN black hole is the limit of the zero-angular momentum $(a = 0)$.
The charged scalar field has the radial part
\begin{eqnarray} \label{ex-bh rad}
\frac{d}{dr} \Bigl((r^2 - B^2) \frac{dR}{dr} \Bigr) + \Bigl[ \frac{\bigl( \omega - \kappa r \bigr)^2}{r^2 - B^2} - \bigl( m^2 l_\mathrm{AdS}^2 + \lambda \bigr) \Bigr] R = 0,
\end{eqnarray}
where
\begin{eqnarray}
\kappa = \frac{(q Q + p P) (Q^2 + P^2) - 2 n a r_\mathrm{H}}{l_\mathrm{AdS}^2},
\end{eqnarray}
and the separation constant $\lambda$ is an eigenvalue of the angular part in the metric (\ref{ex-bh met}). The radial equation has solutions in terms of the hypergeometric function, and imposing the boundary condition on tunneling wave functions in Sec.~\ref{sec2}, the vacuum persistence amplitude and the mean number of pairs are found\cite{Chen:2016caa,Chen:2017mnm,Chen:2019rtd}
\begin{eqnarray} \label{ex-bh al}
|\alpha|^2 &=& \frac{\cosh(\pi \kappa - \pi \mu) \cosh(\pi \tilde\kappa + \pi \mu)}{\cosh(\pi \kappa + \pi \mu) \cosh(\pi \tilde\kappa - \pi \mu)},
\\ \label{ex-bh bet}
|\beta|^2 &=& \frac{\sinh(2 \pi \mu) \sinh(\pi \tilde\kappa - \pi \kappa)}{\cosh(\pi \kappa + \pi \mu) \cosh(\pi \tilde\kappa - \pi \mu)},
\end{eqnarray}
where
\begin{eqnarray} \label{ex-bh par}
\tilde \kappa = \frac{\omega}{B}, \quad \mu = \sqrt{\kappa^2 - m^2 l_\mathrm{AdS}^2 - \lambda - \frac{1}{4}}.
\end{eqnarray}
The violation of the BF bound for spontaneous pair production requires the positivity of $\mu$.

Finally, we comment on the Schwinger effect from the charged KN black holes. First, the leading Boltzmann factor can be obtained from the phase-integral $R= e^{i {\cal S}/[\hbar]}$, which has simple three poles in the complex plane of $r$: $r = \pm B$ and $r = \infty$. The contour clockwise integral enclosing the three poles
\begin{eqnarray}
\Im {\cal S} = \oint \frac{dr}{r^2 - B^2} \sqrt{(\omega - \kappa r)^2 - {\bar m}^2l_\mathrm{AdS}^2 (r^2 - B^2)},
\end{eqnarray}
gives the leading term of the mean number
\begin{eqnarray}
{\cal N} = |\beta|^2 \simeq e^{- 2 \pi (\mu - \kappa)}.
\end{eqnarray}
The Boltzmann factor may be interpreted in terms of the effective temperature in Sec.~\ref{sec3}. The ${\rm AdS}_2$ space binds pairs and increases the effective mass
\begin{eqnarray}
{\bar m} = m \sqrt{1 + \frac{\lambda + 1/4}{m^2 l_\mathrm{AdS}^2}}.
\end{eqnarray}
The effective temperature has the same form as Eq.~({\ref{eff tem})
\begin{eqnarray}
T_\mathrm{eff} = T_\mathrm{U} + \sqrt{T^2_\mathrm{U} + \frac{R_\mathrm{(A)dS}}{2 (2 \pi)^2} }, \label{ex-bh tem}
\end{eqnarray}
where  charges accelerate  and have the Unruh temperature
\begin{eqnarray}
a_{c}= \frac{\kappa}{{\bar m} l_\mathrm{AdS}^2}, \quad T_\mathrm{U} = \frac{a_c}{(2 \pi)}.
\end{eqnarray}
Then, the Schwinger emission has ${\cal N} \simeq e^{- \frac{\bar m}{T_\mathrm{eff}}}$.
Introducing another temperature parameter associated to Eq.~(\ref{ex-bh tem})
\begin{eqnarray}
\tilde{T}_\mathrm{eff} = T_\mathrm{U} - \sqrt{T^2_\mathrm{U} + \frac{R_\mathrm{(A)dS}}{2 (2 \pi)^2} },
\end{eqnarray}
the exact mean number of emitted charges from the near-extremal KN black hole can be factorized as\cite{Kim:2015kna,Chen:2017mnm,Chen:2019rtd}
\begin{equation} \label{ex-bh ther}
\mathcal{N} =  \underbrace{\Biggl( \frac{\mathrm{e}^{-\frac{\bar{m}}{T_\mathrm{eff}}} - \mathrm{e}^{-\frac{\bar{m}}{\tilde{T}_\mathrm{eff}}}}{1 + \mathrm{e}^{-\frac{\bar{m}}{\tilde{T}_\mathrm{eff}}}}\Biggr) }_{\text{Schwinger effect in AdS$_2$}} \times \mathrm{e}^{\frac{\bar{m}}{T_\mathrm{eff}}} \underbrace{\Biggl\{ \frac{\mathrm{e}^{-\frac{\bar{m}}{T_\mathrm{eff}}} \left( 1 - \mathrm{e}^{-\frac{\omega - q \Phi_\mathrm{H} -p \bar\Phi_\mathrm{H} - n \Omega_\mathrm{H}}{T_\mathrm{H}}} \right)}{1 + \mathrm{e}^{-\frac{\omega - q \Phi_\mathrm{H} - p \bar\Phi_\mathrm{H} - n \Omega_\mathrm{H}}{T_\mathrm{H}}} \mathrm{e}^{- \frac{\bar{m}}{T_\mathrm{eff}}}} \Biggr\} }_{\text{Schwinger effect in Rindler$_2$}}.
\end{equation}
The factorization (\ref{ex-bh ther}) is expected for any near-extremal charged black hole because the emission is essentially prescribed by two temperatures: the effective temperature $T_\mathrm{eff}$ and the Hawking temperature $T_\mathrm{H}$. The Schwinger effect has the same effective temperature $T_\mathrm{eff}$ in the $\mathrm{AdS}_2$ space.\cite{Cai:2014qba} The deviation from the extremality measured by $B$ is now given by the Hawking temperature (\ref{ex-bh tem-en}), and  Hawking radiation, though exponentially suppressed, still resides and can be expressed as the Unruh effect in a Rindler space. The mean number and vacuum persistence amplitude for scalar charges in Rindler space was extensively studied in Ref.~\citenum{Gabriel:1999yz}. The charges from the event horizon have the energy $\omega$, the electrical chemical potential $\Phi_\mathrm{H}$, and the magnetic chemical potential $\bar\Phi_\mathrm{H}$ as well as $\Omega_\mathrm{H}$. The emission of charges during the gravitational collapse to an extremal black hole exhibits a nonthermal spectrum.\cite{Gao:2002kz}

In general, $T_\mathrm{eff} \gg T_\mathrm{H}$ for near-extremal black holes, so the Schwinger effect is the dominant channel for charge emission from the black holes. Further, in the extremal limit, the Hawking temperature vanishes, the second factor becomes unity, and the emission of charges is entirely due to the Schwinger effect. It is worth mentioning that the first law of thermodynamics with the effective temperature recovers the entropy of extremal RN black hole provided that $(q = m)$.\cite{Kim:2015qma}

\section{QED Phenomena in Highly Magnetized Stars and Magnetized Black Holes}\label{sec5}

Compact stars, such as neutron stars and magnetars, have strong EM fields beyond the attainability of laboratory experiments possibly except for pulsed and localized EM fields from ultra-intense lasers, as will be discussed in Sec. \ref{sec7}. Many of neutron stars have surface magnetic fields comparable to the critical field intensity and magnetars, the highly magnetized neutron stars, have magnetic fields two order higher or beyond the critical value $B_c$.\cite{Kaspi:2017fwg} The virial theorem of magnetohydrodynamic equilibrium, which prescribes the total energy of magnetic field to be no greater than the gravitational potential energy, constrains the upper limit to the magnetic field strength of a neutron star \cite{Shapiro:1983}
\begin{eqnarray}
B_b \leq 10^{18} \Bigl(\frac{M_{\rm ns}}{M_{\odot}} \Bigr) \Bigl(\frac{10\,km}{R_{\rm rn}} \Bigr)^2\,G.
\end{eqnarray}
The dynamo processes in proto-neutron stars can generate magnetic fields of order $10^{15}\, {\rm G}$ or stronger,\cite{Duncan:1992hi} and magnetars were observed.\cite{Vasisht:1997je} Strong EM fields not only drastically change quantum states of matter and the equation of state different from those of the ordinary matter in weak EM fields, but also modify the propagation and spectrum of photons from the Maxwell theory. The state of matter (atoms) changes when the Landau level energy in a magnetic field equals to the Bohr atomic energy, which gives a field strength $B_0 = [c] e^3 m^2/[\hbar^3] = 2.35 \times 10^9\, \mathrm{G}$,.\cite{Meszaros:1992} The EM field can change the vacuum structure of the Dirac sea when the cyclotron energy equals to the rest mass energy of electron, which gives the critical field strength $B_c = [c^2]m^2/[\hbar] e = 4.41 \times 10^{13}\, \mathrm{ G}$, or the potential energy over one Compton wavelength of electron equals to the rest mass energy $E_c = [c^2]m^2/[\hbar] e = 1.32 \times 10^{16}\, \mathrm{ V/cm}$.\cite{Dittrich:1985yb,Greiner:1985ce}

The effective action (\ref{qed gen}) of the Maxwell term and the one-loop QED action leads to the induced fields
\begin{eqnarray}
{\bf D} = 4 \pi \frac{\delta {\cal L}_{\rm eff}}{\delta {\bf E}}, \quad {\bf H} = - 4 \pi \frac{\delta {\cal L}_{\rm eff}}{\delta {\bf B}}.
\end{eqnarray}
Up to the leading order of the one-loop effective action the dielectric and permeability tensors ($D_i = \epsilon_{ij} E_j$ and $H_i = \mu^{-1}_{ij} B_j$) are given by\cite{Ruffini:2009hg}
\begin{eqnarray}
\epsilon_{ij} = \delta_{ij} + \frac{1}{45 \pi} \Bigl(\frac{e}{m}\Bigr)^4 \Bigl[ 2 \bigl( {\bf E}^2 - {\bf B}^2 \bigr) \delta_{ij}
+ 7 B_i B_j \Bigr], \nonumber\\
\mu^{-1}_{ij} = \delta_{ij} + \frac{1}{45 \pi} \Bigl(\frac{e}{m}\Bigr)^4 \Bigl[ 2 \bigl( {\bf E}^2 - {\bf B}^2 \bigr) \delta_{ij}
- 7 E_i E_j \Bigr], \label{biref}
\end{eqnarray}
where the one-loop contributions are order of $(\alpha = e^2/4 \pi)^2$. Then, probing photons passing through a strong magnetic field experience the vacuum birefringence of $n_{\perp}$ and $n_{\parallel}$, depending on the propagation directions.

The vacuum birefringence has not been measured at terrestrial experiments yet (for recent review, see Ref.~\citenum{Zavattini:2018pbw}). The photons emitted from the surface of a neutron star experience two different refringences and result in the polarization effect,\cite{Heyl:2002vc} which has been recently observed by optical polarization measurements from  RX J1856.5−3754, an isolated neutron star, with $10^{13}\, {\rm G}$.\cite{Mignani:2016fwz} Another QED effect in strong EM fields will be the photon splitting from highly magnetized stars. The Furry theorem prohibits any scattering matrix of odd number of external lines and photons. The lowest nonvanishing Feynman loop diagram has 3 vertices of external fields and 3 vertices of photons, which predicts one photon to split into two photons.\cite{Adler:1971wn} The QED physics in highly magnetized neutron stars is discussed in Ref. \citenum{Harding:2006qn}, the nuclear and quark matters in Ref.  \citenum{Lattimer:2015nhk}, and probing QED effects is proposed using MeV telescope.\cite{Wadiasingh:2019jdr}

Another astrophysical sources of EM fields are black holes, which can be formed from binary neutron star mergers\cite{TheLIGOScientific:2017qsa} or gravitational collapse of unstable magnetars.\cite{Falcke:2013xpa} One may question the fate of strong magnetic fields after black holes are formed from mergers or collapses. In the general relativity, a few solutions of the Einstein-Maxwell equation were found. The Bonnor-Melvin universe or magnetized universe has a uniform magnetic field even in the asymptotic region. \cite{Bonnor:1954,Melvin:1964} Though the magnetic universe may have a cosmological interest in relation with the anisotropy of CMB, the astrophysical objects with magnetic fields extending to the spatial infinity cannot be physically plausible except for approximate solutions.

The charged nonrotating RN black hole can emit charges through the Schwinger mechanism.\cite{Zaumen:1974,Carter:1974yx,Gibbons:1975kk,Damour:1976jd} Damour and Ruffini studied the Schwinger mechanism in the charged rotating KN black hole,\cite{Damour:1974qv} which led Ruffini and his collaborators to propose the dyadosphere model of RN black holes for GRBs\cite{Ruffini:1998,Preparata:1998rz} and later on dyadotorus model of KN black holes.\cite{Cherubini:2008zz,Cherubini:2009ww} However, Page refuted the dyadosphere model because the enormous disparity between the strong electrostatic repulsion and the weak gravitational attraction of charged particles prohibits the formation of a charged nonrotating RN black hole from a gravitational collapse.\cite{Page:2006cm} A similar conclusion was reached for a charged black hole in the $\mathrm{dS}$ space.\cite{Moradi:2017alp} This does not completely rule out the black hole model with magnetic fields.\cite{Cherubini:2018fba}

The dyonic KN black hole has both electric and magnetic fields associated to their electric and magnetic charges. Without magnetic monopoles, this black hole cannot properly describe an astrophysical black hole with a magnetic field. Wald showed that the Killing vector of a rotating black hole gives the vector potential of an external magnetic field around the black hole, which allows accretion of charges into the black hole.\cite{Wald:1974np} The magnetic field of arbitrary strength can exist around a Kerr black hole and accredit sufficient amount of charges.\cite{Dokuchaev:1987ova}
The magnetized black holes have since then been studied in connection with astrophysics.\cite{Aliev:1989wx} The magnetized electric RN black hole and KN black hole as the solutions of the Einstein-Maxwell theory were obtained in Ref. \citenum{Gibbons:2013yq}. These solutions, which include the Wald solution in the weak field limit, however, still have asymptotic EM fields. No theorem has been proved to prohibit black holes from carrying localized magnetic fields. If such magnetic black holes exist, they may explain the high energy cosmic rays and support the BZ mechanism for GRBs. The Meissner effect expelling magnetic field from extremal KN black holes is a dilemma in extracting the rotational energy from the holes,\cite{King:1975tt} and has recently been discussed.\cite{Penna:2014aza}

\section{QED Phenomena in the Early Universe and Magnetogenesis}\label{sec6}

The physical motivation to study QED phenomena in the early universe, not to mention a theoretical interest, is the magnetogenesis and the electroweak phase transition, which are believed to occur in some stage of the evolution of the universe. The magnetogenesis, the origin of cosmic scale magnetic fields with the current bound $B_c > 10^{-15} \, \mathrm{G}$,\cite{Neronov:1900zz} is an open question in the early universe (for a comprehensive review and references, see Ref.~\citenum{Durrer:2013pga}). Various scenarios have been proposed for the magnetogenesis, such as phase transitions during the inflation.\cite{Durrer:2013pga} One scenario is the Schwinger mechanism for inducing a current during the inflation epoch because even an extremely weak electric field can generate the current greatly enhanced by Hawking radiation with large Hubble constant and high effective temperature,\cite{Bavarsad:2017oyv} as explained in Sec.~\ref{sec3}. Then, the induced current generates a magnetic field at the cosmic scale whose flux without plasma interactions would be conserved and the magnetic field in turn amplifies the induced current which is proportional to the magnetic field itself in the density of states. This self-accelerating generation of magnetic fields through the Schwinger mechanism may be an origin for the magnetogenesis.\cite{Kobayashi:2014zza,Bavarsad:2017oyv,Sharma:2017eps,Stahl:2018idd,Sobol:2018djj,Kobayashi:2019uqs} The IR-hyperconductivity may play some role for amplifying the magnetic fields.\cite{Frob:2014zka,Bavarsad:2016cxh,Kobayashi:2014zza,Hayashinaka:2016qqn,Hayashinaka:2016dnt}

The Schwinger effect in the $\mathrm{dS}$ space or during the inflation period is both a theoretically and physically interesting issue since the $\mathrm{dS}$ space drives quantum fields into a thermal state and emits Hawking radiation with Gibbons-Hawking temperature.\cite{Gibbons:1977mu} Hawking radiation is an effect of nonperturbative quantum fields. The issues in the dS space, such as the stability and density perturbations for cosmic microwave background radiation etc, will not be discussed in this review. As explained in Sec.~\ref{sec3}, the Schwinger effect is another prediction of nonperturbative quantum field in EM fields. It is therefore interesting and challenging to study the Schwinger effect in the $\mathrm{dS}$ space.

The Schwinger effect in the $\mathrm{dS}$ space has various motivations, such as bubble nucleation in the early universe, the IR-hyperconductivity, the vacuum polarization and pair production of scalars in the $\mathrm{dS}_2$ space,\cite{Cai:2014qba,Frob:2014zka} and the equivalence between fermions and bosons rate in $\mathrm{dS}$ space.\cite{Stahl:2015gaa} The dimensional dependence of Hawking in the global coordinates of dS space, which can be explained as the Stokes phenomena of constructive or destructive interference,\cite{Kim:2010xm,Kim:2013cka,Dabrowski:2014ica} motivated the Schwinger effect of scalars in the ${\rm dS}_3$ space,
which showed no destructive interference in the presence of an electric field.\cite{Bavarsad:2016cxh} The mean number of scalar pairs and the classicality parameter of the position and momentum operator correlation were numerically computed,\cite{Sharma:2017ivh} and the coupling of the Maxwell theory to an inflaton gave the power-law behavior of the induced current on the electric field.\cite{Geng:2017zad}

The issue of controversial negative conductivity was studied in detail in Ref.~\citenum{Banyeres:2018aax}. The Schwinger subtraction scheme for renormalizing the one-loop effective action together with the Schwinger effect in ${\rm dS}_2$ space, as shown in Sec.~\ref{sec3},\cite{Bavarsad:2017oyv} suggests that the negative conductivity may be an artifact of the renormalization scheme. The IR-hyperconductivity, however, has been confirmed in different renormalization schemes: the Pauli-Villars method,\cite{Frob:2014zka,Banyeres:2018aax} the adiabatic substraction method,\cite{Stahl:2015gaa,Bavarsad:2016cxh,Kobayashi:2014zza,Hayashinaka:2016qqn} and the point-splitting.\cite{Hayashinaka:2016dnt} It was argued that the maximal subtraction scheme removes the IR-hyperconductivity.\cite{Hayashinaka:2018amz} In the adiabatic substraction renormalization scheme, the vector potential of the first adiabatic order instead of the zeroth recovers the know results of the trace and chiral anomaly.\cite{Ferreiro:2018qdi,Ferreiro:2018qzr}.

The backreaction problem of the produced pairs through the Schwinger mechanism together with Hawking radiation has been known as an interesting physical issue requiring a further study. In the Minkowski spacetime, the created pairs induce a current which increases in time because the pairs are continuously created and added to the existing current. The pairs are spontaneously produced per unit four-volume at the cost of the field and generate the induced current increasing in time. Then, the increasing induced current generates a magnetic field increasing in time, which in turn induces a counter electric field in opposite direction to the original field. The overshooting of the counter electric field makes the total electric field and plasma oscillating.\cite{Kluger:1991ib} The backreaction of the Schwinger effect in the dS space discussed in Refs.~\citenum{Kobayashi:2014zza,Sharma:2017eps,Sobol:2018djj,Banyeres:2018aax} and \citenum{Stahl:2016geq} decays both the background electric field and the Hubble constant of the vacuum energy.\cite{Bavarsad:2019cd} It was argued that the backreaction of the Schwinger effect may violate the second law of thermodynamics.\cite{Giovannini:2018qbq} To sustain the electric field, an anisotropic inflation and a dilaton coupling of the Maxwell theory were proposed,\cite{Kitamoto:2018htg} but a more complete model may not allow for a large constant electric field in the dS space.\cite{Shakeri:2019mnt}

Not considered in this review are the $\mathrm{ER=EPR}$ conjecture with holographic Schwinger effects, \cite{Fischler:2014ama} and the Schwinger mechanism in SU(2) field,\cite{Lozanov:2018kpk,Maleknejad:2018nxz} Hawking radiation and Schwinger effect during the electroweak phase transition, and cosmological implications to the gauge-flation\cite{Maleknejad:2012fw} and the magnetogenesis. In order for the Schwinger mechanism to efficiently act to induce current, a complete model is necessary because the backreaction of the induced current decays both the background field and the Hubble constant.

\section{Laboratory Astrophysics using Ultra-intense Lasers}\label{sec7}

In astrophysics strong EM fields are important ingredients in addition to strong gravitational fields. The gravity effect of compact stars, such as neutron stars or black holes, can be simply described in terms of physical parameters as shown in Table~\ref{aba:tbl3}. Strong gravitational fields originate from huge masses compacted into small regions due to the small Newton constant. Such high density matter cannot be realized in  laboratory at the macroscopic scale. The equivalence principle of general relativity, however, provides us with an alternative: the acceleration of detectors. The nonperturbative quantum effect, such as Hawking radiation of a black hole, can also be explained as the Unruh effect, whose temperature is proportional to the acceleration of an observer at the horizon of black hole. This implies that when one can accelerate the system under study, the system emulates environments with strong gravity and EM fields.

In Table~\ref{aba:tbl4} the EM field intensity, acceleration, temperature, and mass (energy) density achievable in laboratory are compared with those of astrophysical sources. 	In laboratory, those extreme physical values can be obtained with ultra-intense lasers. Based on the chirped pulse amplification technique,\cite{Strickland:1985} the current ultra-intense lasers can deliver pulses having a duration of a few tens of femtoseconds and an energy of over $100\, \mathrm{J/pulse}$.\cite{danson15} When such a laser pulse is focused to its physical limit, i.e. its wavelength ($\sim\mu \mathrm{m}$), the intensity can be as high as $10^{23}\, \mathrm{W/cm^2}$: when a $10\, \mathrm{fs}$, $100\,\mathrm{J/pulse}$ is focused on an area of $10 \, \mu\mathrm{m^2}$, an intensity of $10^{23}\, \mathrm{W/cm^2}$ is achieved. The current intensity record is $5.5\times10^{22} \, \mathrm{W/cm^2}$ obtained with the 4-PW laser at CoReLS,\cite{jhsung17,jwyoon19} and several laser mega-projects are ongoing for building more powerful lasers and using them for fundamental physics and applications, e.g., Extreme Light Infrastructure-Nuclear Physics in Romania (10-PW lasers),\cite{elinp} and Station of Extreme Light in China (100-PW laser).\cite{shen18} Even with a 100-PW laser, it is difficult to obtain an intensity over $10^{24} \, \mathrm{W/cm^2}$, and a scheme based on a laser-driven relativistic flying mirror was proposed to reach an intensity close to the critical intensity for the vacuum breakdown ($\sim 10^{29} \, \mathrm{W/cm^2}$); in the scheme the laser wavelength is shortened by Doppler effect to the x-ray region.\cite{bulanov11}

\begin{table}
\tbl{Comparison of Astrophysics and Laboratory Astrophysics}
{\begin{tabular}{@{}cccc@{}}
\toprule
Astrophysics  &  Unruh temperature & Acceleration mechanism & Electromagnetic Fields \\
vs Laboratory & & & \\
\hline\\[-6pt]
Black holes & $T_{\rm H} = \frac{\kappa[\hbar]}{2 \pi [c]} $ & surface gravity $\kappa$  & black hole charges or disks  \\
Neutron stars & $T_{\rm U} = \frac{a[\hbar]}{2 \pi [c]} $ & gravitational acceleration & polar magnetic fields\\
\hline\\[-6pt]
Laboratory & $T_{\rm U} = \frac{a[\hbar]}{2 \pi [c]} $ & vacuum laser acceleration & ultra-intense laser fields \\
 &   & laser wakefield acceleration & laser-induced plasma fields \\
\botrule
\end{tabular}\label{aba:tbl3}
}
\end{table}

\begin{table}
\tbl{Current Status of Laboratory Experiments$^{\text a}$}
{\begin{tabular}{@{}ccc@{}}
\toprule
Physical parameters  &  Laboratory & Compact stars \\
\hline\\[-6pt]
EM intensity & ELI-NP$^{\text b}$  & Neutron stars/magnestars  \\
& $E = 10^{-3} E_c$ &  $B_\mathrm{ns} = 10^{-5} \thicksim 10^{2} B_c$ \\
&IZEST$^{\text c}$ &\\
& &\\
\hline\\[-6pt]
Acceleration & Linear acceleration $(m/s^2)$ $^{\text d}$  & Black holes/Neutron stars $(m/s^2)$\\
& Optical pulse $a= 10^{23} $ &   $\kappa = 1.53 \times 10^{13} \Bigl(\frac{M_{\odot}}{M_\mathrm{bh}} \Bigr)$\\
&X-ray pulse $a= 10^{26} $ & $g_{ns} = 1.33 \times 10^{12} \Bigl(\frac{M_\mathrm{ns}}{M_{\odot}} \Bigr) \Bigl(\frac{R_{10}}{R_\mathrm{ns}} \Bigr)^2 $\\
\hline\\[-6pt]
Temperature & $T_{\rm U} = 3.50 \times \frac{a (m/s^2)}{10^{25}}~(eV)$ & $T_\mathrm{H} = 5.32 \times 10^{-12} \Bigl(\frac{M_{\odot}}{M_\mathrm{bh}} \Bigr)~(eV)$\\
\hline\\[-6pt]
Mass(energy) density & RHIC$^{\text f}$ & Black holes$^{\text e}$/Neutron stars\\
&$\epsilon = 1.8 \times 10^{16} (g/cm^3)$ & $\rho_{\rm Sch} = 1.84 \times 10^{16} \Bigl(\frac{M_{\odot}}{M_\mathrm{bh}} \Bigr)^2 (g/cm^3)$ \\
& $T = 2.85 \times 10^{8} (eV)$  & $ \rho_\mathrm{ns} = 3.7 \thicksim 5.9 \times 10^{14} (g/cm^3)$ \\
\botrule
\end{tabular}
}
\begin{tabnote}
$^{\text a}$ Electromagnetic fields are in expressed in terms of critical strength $E_c = 1.32 \times 10^{16}~ (V/cm)$, $B_c = 4.41 \times 10^{13}~ (Gauss)$, and intensity $I_c = 2.3 \times 10^{29}~(W/cm^2)$.\\
$^{\text b}$ Extreme Light Infrastructure-Nuclear Physics.\\
$^{\text c}$ International Zeta-Exa Watt Science and Technology.\\
$^{\text d}$ Acceleration of charge in an intense laser $a < \frac{c}{t_{\rm optical\, period}} = \frac{c^2}{\lambda}$. \\
$^{\text e}$ Schwarzschild density $\rho_{\rm Sch} = \frac{3 c^6}{32 \pi G^3 M_{\rm bh}^2}$.\\
$^{\text f}$ Relativistic Heavy Ion Collider.
\end{tabnote}\label{aba:tbl4}
\end{table}

With such ultra-intense lasers, an extreme value of acceleration can be realized in laboratory. Upon colliding with a relativistic electron, an ultra-intense laser pulse can drive the electron in the opposite direction up to the speed of light within a fraction of an optical period.\cite{Kim:2017ets} A rough estimation of the acceleration is given as $a=c/(\lambda/c)$, which is about $10^{23} \,\mathrm{m/s^2}$ for the laser wavelength $\lambda\sim 1 \, \mu\mathrm{m}$. The corresponding Unruh temperature is $3.5 \times 10^{-2} \mathrm{eV}$. If the wavelength is shortened to the x-ray region ($\lambda\sim 10^{-3} \mu\mathrm{m}$), e.g., by the flying mirror scheme, the Unruh temperature will increase to $35 \, \mathrm{eV}$ ($\sim 4 \times 10^5 \, \mathrm{K}$), which should be measurable. One conceptual and technical question is how to keep the acceleration phase long enough to guarantee the Unruh effect, which is a consequence of the infinite duration of uniform acceleration. One may consider a circular motion of electrons using ring accelerators, but the centripetal acceleration is not uniform, whose Unruh effect was debated.\cite{Unruh:1998gq} The energy density of a circularly rotating detector from the Wightman function shows a different spectrum from that of a uniformly accelerating detector.\cite{Kim:1987pv}

Considering the rapid development of ultra-intense lasers, strong field QED effects are likely to be observed in the near future. As stated in Sec.~2, the QED physics in strong EM fields are related to either the vacuum polarization or the vacuum persistence amplitude. The latter is a consequence of the Schwinger effect as shown in Eq.~ (\ref{vacuum persistence}). In the case of a constant electric field parallel to a constant magnetic field, the vacuum polarization is given by Eq.~(\ref{qed E-B}) and the vacuum persistence amplitude by Eq.~(\ref{vac E-B}). The QED action in a constant EM field has been theoretical interest since the seminal works by Heisenberg-Euler and Schwinger.\cite{Heisenberg:1935qt,Schwinger:1951nm}

In the experiments using ultra-intense lasers, the EM fields have nontrivial configurations. Finding explicitly the QED actions in pulsed and/or localized EM fields has been a challenging task since Heisenberg, Euler and Schwinger. The resolvent operator method was used to find analytical expressions for QED action in the Sauter-type electric field, $E(t) = E_0/\cosh^2 (t/T)$.\cite{Dunne:1998ni} Recently, the gamma-function regularization has been introduced to calculate QED actions in the Sauter-type electric and magnetic fields. The gamma-function regularization turns out useful when the exact solutions of the field equation are known in a given EM field and thereby the Bogoliubov relation is explicitly given in terms of gamma functions. To the best knowledge of the author, most exactly solved field equations exhibit this property. In the generic case, in which the solutions are not known in terms of special functions, the numerical methods based on the worldline path integral or the Wigner formalism seem to be an option.

On the other hand, various methods have been employed to find the Schwinger effect. Two of the most widely used methods are the worldline instanton method\cite{Dunne:2005sx,Dunne:2006st} and the phase-integral method.\cite{Kim:2000un,Kim:2003qp,Kim:2007pm} The momentum integral of the phase-integral up through quadratic order gives the worldline instanton result including the prefactor for a given profile of electric field depending on one time or space coordinate.\cite{Kim:2019yts} The worldline path integral and the Wigner formalism have been used for numerical calculations of pair production in general profiles of electric fields. Provided that the mean number of produced pairs can be exactly found, the vacuum persistence amplitude (\ref{vacuum persistence}) may reconstruct the vacuum polarization through the Mittag-Leffler theorem and the renormalization of the vacuum energy and charge, which is a one-loop action analog of the optical theorem in particle physics.\cite{Kim:2016nyz}  The mean number from various methods can give an approximation of the vacuum polarization; for instance, the Liouville-Green function method gives the leading pair production and the corresponding vacuum polarization, which confirms the reconstruction conjecture from the vacuum persistence amplitude.\cite{Kim:2009pg}

\section{Conclusion}

It has passed over one hundred years since Einstein discovered the general relativity and over eighty years after Heisenberg and Euler found the QED one-loop effective action in a constant electromagnetic field. Einstein gravity differs from the Newtonian gravity in strong gravitating systems and cosmology, and the Heisenberg-Euler and Schwinger QED action in a strong electromagnetic field modifies the Maxwell theory into a nonlinear theory with the vacuum polarization due the interaction of photons with virtual electrons from the Dirac sea. Further, a strong electric field can spontaneously create electron-positron pairs, the (Sauter-)Schwinger pair production, from the Dirac sea. The noticeable effects of general relativity and the QED action occur when the gravity and EM fields are strong. In cosmos, strong gravitating objects, such as compact stars and black holes, and the inflation model are best described by the general relativity. The vacuum birefringence effect of QED action has been measured by observing the optical spectrum from a neutron star. With the recent observations of gravitational waves and the development of ultra-intense lasers using CPA technology, the physics of strong gravity and EM fields are emerging as a new window to probe the fundamental nature of the gravitational and/or electromagnetic interactions and the quantum vacuum structure.

In this review, we critically review QED actions in strong electromagnetic fields in the Minkowski spacetime or curved spacetimes. One physical motivation for studying QED phenomena in strong electromagnetic fields is the recent development of ultra-intense lasers using CPA technology, which have already exceeded the relativistic region and is rapidly approaching toward the critical field strength of electron-positron pair production. Another motivation comes from astrophysical objects, such as highly magnetized neutron stars and, in particular, magnetars with the surface field strength two order higher and much higher at the core than the critical field. The energy density of the magnetic field on the surface of magnetars is roughly thousand times higher than the electron energy density per unit Compton volume and much higher at the core. The electromagnetic interactions between charged particles and photons in such strong magnetic fields differ from those without magnetic fields. Hence, the physical model for neutron stars or magnetars and their binary mergers or gravitational collapses should necessarily include the QED effects. A (near-)extremal black hole has a tiny Hawking temperature and Hawking radiation is exponentially suppressed. The near-horizon geometry of the (near-)extremal black hole is a warped $\mathrm{AdS}$ space, whose symmetry allows explicit solutions of charged fields and the pair production.

Still another interesting arena for strong electromagnetic fields is the universe. The cosmic scale magnetic field of $\mathrm{nG}$, without plasma dissipation mechanism, would amplify to an enormous strength in the early universe due to the flux conservation. In this scenario for the cosmic magnetic field, the electrodynamics should be properly described by the Maxwell-QED theory. Recently, the Schwinger mechanism has been studied in de Sitter spaces. The induced current by an electric field in the $\mathrm{dS}$ consists of two parts: the Schwinger effect which dominates in strong field limit, and Hawking radiation whose charged pairs are aligned by the field in the weak field limit. The Schwinger effect and Hawking radiation are intertwined in general as manifested by the effective temperature of the geometric mean of the Unruh temperature and the Gibbons-Hawking temperature. In the very weak field limit, charges from Hawking radiation provide a persistent current and lead to the IR-hyperconductivity. The Schwinger mechanism has been proposed for magnetogenesis.

\section*{Acknowledgments}
 The author is deeply indebted to Chul Min Kim and Cl\'{e}ment Stahl in preparing laboratory astrophysics and  QED in dS space and magnetogenesis, particularly grateful to Rong-Gen Cai for discussions on the effective temperature in high dimensions, and Chiang-Mei Chen for discussions on black holes. For collaborations and useful discussions, he also  would like to thank Fiorenzo Bastianelli, Ehsan Bavasard, Rong-Gen Cai, Chiang-Mei Chen, Alexei Gaina, Gary Gibbons, Mohammad Ali Gorgi, Takahiro Hayashinaka, Hyun Kyu Lee, Hyung Mok Lee, Chang Hee Nam, Wei-Tou Ni, Don N. Page, Miok Park, Christopher N. Pope, Remo Ruffini, Misao Sasaki, Christian Schubert, Jia-Rui Sun, She-Sheng Xue, Jun'ichi Yokoyama, and Yongsung Yoon. He appreciates the warm hospitality at Institute of Theoretical Physics (ITP), Chinese Academy of Sciences (CAS) during his sabbatical leave. This work was supported in part by Institute for Basic Science (IBS) under IBS-R012-D1, Korea, and by National Research Foundation of Korea (NRF) funded by the Ministry of Education (NRF-2015R1D1A1A01060626) and in part by the Open Project Program of State Key Laboratory of Theoretical Physics, ITP, CAS, China (No.~Y5KF161CJ1) and the Fundamental Research Funds for the Central Universities.

%\bibliographystyle{ws-procs961x669}
%\bibliography{ws-pro-sample}

%\end{document}

%Non BiBTeX users can list down their references as:

\end{document}